\documentclass[preprint,showpacs,preprintnumbers,amsmath,amssymb]{revtex4}

\usepackage[dvips]{graphicx}
\usepackage{dcolumn}
\usepackage{bm}

  \newcommand{\nn}{\nonumber}

  \newcommand{\sura}{\hspace{-1.5mm}\!/}
  
\begin{document}

\title{Spin analyzing power for polarized top decays with jets}

\author{Yoshio Kitadono$^1$}
 \email{kitadono@phys.sinica.edu.tw}
\author{Hsiang-nan Li$^{1,2,3}$}
 \email{hnli@phys.sinica.edu.tw}
 \affiliation{$^1$Institute of Physics, Academia Sinica,
 128 Sec.2, Academia Rd., Nankang, Taipei 11529, Taiwan, Republic of China}
 \affiliation{$^2$Department of Physics, National Cheng-Kung university,
 Tainan, Taiwan701, Republic of China}
 \affiliation{$^3$Department of Physics, National Tsing-Hua university,
 Hsin-Chu, Taiwan300, Republic of China}


\date{\today}

\begin{abstract}
We perform perturbative QCD factorization of infrared radiations
associated with an energetic $b$ quark from a polarized
top quark decay, taking the semi-leptonic channel as an example.
The resultant formula is expressed as a convolution of an infrared-finite heavy-quark
kernel with a $b$-quark jet function. Evaluating the heavy-quark
kernel up to leading order in the coupling constant and adopting the jet
function from QCD resummation, we predict the dependence of the spin
analyzing power for a polarized top quark on the invariant mass of
the $b$-quark jet. It is observed
that the spin analyzing power could be enhanced by a factor 2
compared to the inclusive case with the jet mass being
integrated over. It is worthwhile to test experimentally
the enhancement of the spin analyzing power due to
the inclusion of jet dynamics.
\end{abstract}



\maketitle

\section{Introduction}

A study of the top quark physics contributes to the understanding of
the origin of the electro-weak symmetry breaking
in the standard model and its extensions. Progress in this field
from the Large Hadron Collider (LHC), a top quark factory,
has been reviewed in~\cite{Topreview1,Topreview2}. Higher-order calculations
of top production and decay processes were summarized
in~\cite{TopreviewNLO} (see also references therein). Especially,
observables related to the spin of a top quark may reveal
various new physics effects, which can be explored at LHC (for recent
references, refer to \cite{New}). Since a top quark decays almost $100\%$ into
a $b$ quark and a $W$ boson before hadronization, one can extract the
information on the spin of a "bare top quark" from the angular
distribution of its decay products. For this purpose,
the spin analyzing power $\kappa_{i}$ for a final state $i$
in a polarized top quark decay has been defined via
\begin{eqnarray}
\frac{1}{\Gamma}\frac{d\Gamma}{d\cos\theta_{i}}
= \frac{1}{2}\left( 1 + \kappa_{i}|\vec{P}|\cos\theta_{i} \right),\label{par}
\end{eqnarray}
where $\Gamma$ is the partial decay width, $\vec{P}$ is the polarization
vector of a polarized ensemble of top quarks at rest with $0\le |\vec{P}| \le 1$,
and $\theta_{i}$ is the
angle of the particle momentum measured from the top quark spin.
That is, $\kappa_{i}$ relates the top quark spin to the angular distribution
of the decay product $i$.

QCD corrections to the polarized $t \to bW^{+}$ decay were investigated
in~\cite{TopdecayWNLO}, and those to the spin analyzing power
were done for semi-leptonic
decays in~\cite{TopdecayNLOlep1,TopdecayNLOlep2}, and for hadronic
decays in~\cite{TopdecayNLOhad}. The results have been
summarized in Table 3 of Ref.~\cite{Topreview2}: charged leptons and
down-type quarks exhibit the largest power ($\kappa \simeq 1$),
a bottom quark has a second-largest power ($\kappa \simeq -0.4$), neutrinos
and up-type quarks have a value of $\kappa \simeq -0.3$, and the less energetic
non-$b$ jet shows $\kappa \simeq 0.5$~\cite{less.energetic}. It was observed
that next-to-leading-order (NLO) corrections to the spin analyzing power
are not important, and the $b$ quark mass could be safely ignored
in the above analysis of the polarized top quark decay~\cite{TopdecayNLOhad}.
Although the light-particle jets from the $u$ and $d$ quarks have been considered,
the spin analyzing power associated with the $b$-quark jet is not known yet.
In particular, a jet mass can be measured,
so it is interesting to examine how the spin analyzing power for a polarized
top quark depends on the $b$-quark jet mass. This is the motivation of
the present work.

A fundamental framework for studying high-energy processes is the
perturbative QCD (pQCD) factorization theorem~\cite{Factorization1,Factorization2,Factorization3},
in which cross sections and decay widths are factorized into
convolutions of several subprocesses. Take top quark decays as an example.
The subprocesses include jet functions which
contain infrared radiations associated with energetic hadronic final states,
soft functions which collect infrared gluons exchanged among the top
quark and hadronic final states, and heavy-quark kernels from the
difference between the QCD diagrams for the top quark decays and
the effective diagrams for the jet and soft functions. In this
paper we shall demonstrate the factorization of the $b$-quark jet
from the semi-leptonic top quark decay, that requires
the eikonal approximation for particle propagators in
infrared regions and on the Ward identity for organizing all diagrams
with attachments of infrared gluons \cite{Li01}. The soft functions
will be neglected here due to the cancellation between virtual and real
corrections, and to the fact that they have minor effects
on jet mass distributions. The pQCD factorization
formula for the semi-leptonic decay of a polarized top quark is then
expressed as a convolution of the infrared-finite heavy-quark
kernel with the $b$-quark jet function.

It has been known that
the overlap of collinear and soft dynamics produces large double logarithms
in a jet function, which should be summed up to all orders.
References on the resummation of various double logarithms can be found
in~\cite{Collins.review,Factorization.old1,Factorization.old2,Factorization.old3,Factorization.review}.
Recently, the QCD resummation technique for light-particle jets have been
developed~\cite{Lijetl,Lijet}, by means of which jet substructures,
such as jet mass distributions and energy profiles,
can be calculated. Substituting the quark jet function derived in~\cite{Lijet}
into the pQCD factorization formula for the polarized top quark decay,
we show that the spin analyzing power increases with the jet mass
quickly, and is possibly enhanced by a factor 2 in a wide
range of the jet mass. This result is attributed to the observation that
the shape of the jet function puts more weight on the contribution from
the kinematic region with higher $b$-quark jet momentum, where the
spin analyzing power is larger. It is worthwhile to test experimentally
the enhancement of the spin analyzing power due to
the inclusion of jet dynamics.

The pQCD factorization of the $b$-quark jet function from the semi-leptonic
decay of a polarized top quark is performed up to NLO
by employing the eikonal approximation in Sec.~II. The procedures that rely on the
application of the Ward identity for the
all-order factorization are outlined. The doubly differential width for the
angular distribution of the $b$-quark jet with a specified invariant mass
is presented in Sec.~III. We propose a
parametrization for the quark jet function
derived in \cite{Lijet} in order to simplify numerical analysis. The
choices of physical parameters and our predictions are summarized.
Section~\ref{conclusion} is devoted to a conclusion.

\section{Formalism \label{formalism}}

\begin{center}
 \begin{figure}[htb]
  \includegraphics[scale=0.5]{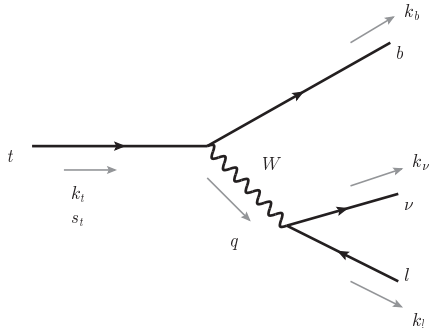}
  \caption{LO amplitude for the semi-leptonic top quark decay.} \label{fig1}
 \end{figure}
\end{center}

We consider the semi-leptonic top quark decay
\begin{eqnarray}
t(k_t, s_t)  &\to& b(k_b) + {\ell}^{+}(k_{\ell}) + \nu_{\ell}(k_{\nu}),
\label{eq.b.parton}
\end{eqnarray}
where $k_t$, $k_{b}$, $k_{\ell}$, and $k_{\nu}$ are the momenta of the top quark,
the $b$ quark, the charged
lepton, and the neutrino, respectively, as indicated in Fig.~\ref{fig1}
for the leading-order (LO) amplitude.
In this section we shall identify the leading infrared contributions,
and perform their factorization into the $b$-quark jet function.

\subsection{Factorization at LO}
\begin{center}
 \begin{figure}[htb]
  \includegraphics[scale=0.5]{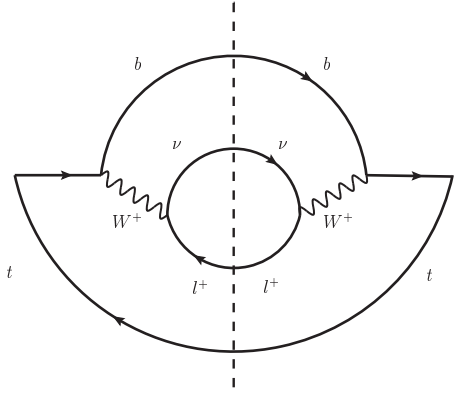}
  \caption{LO decay width for the semi-leptonic top quark decay, where
  the vertical dashed line represents the final-state cut.} \label{fig2}
 \end{figure}
\end{center}

The averaged decay amplitude squared $|\overline{\mathcal{M}}_{0}|^2$ at
LO in both the electro-weak coupling and the strong
coupling, depicted in Fig.~\ref{fig2}, is written as
\begin{eqnarray}
|\overline{\mathcal{M}}_{0}|^2
&=& \frac{1}{4}\frac{ g^4|V_{tb}|^2}{(q^2-m^2_{W})^2
+m^2_{W}\Gamma^2_{W}} L_{\mu\rho} T^{\mu\rho}, \label{lo}
\end{eqnarray}
with the leptonic and hadronic tensors
\begin{eqnarray}
L_{\mu\rho}
&=& d_{\mu\nu}d_{\rho\sigma}
    \mbox{tr}\left[\gamma^{\nu}P_{L}k\sura_{l}
    \gamma^{\sigma}P_{L}k\sura_{\nu}\right],\nn\\
T^{\mu\rho}
&=& \mbox{tr}\left[\gamma^{\mu}P_{L}\hat{w}_{t}(k\sura_{t}+m_{t})
	       \gamma^{\rho}P_{L}(k\sura_{b}+m_{b}) \right]. \label{eq.M0sq}
\end{eqnarray}
In the above expression $g$ is the gauge coupling of the weak interaction,
$V_{tb}$ is the Cabibbo-Kobayashi-Maskawa matrix element,
$q=k_{t}-k_{b}=k_{\nu}+k_{\ell}$, $m_{W}$ and $\Gamma_{W}$ are
the momentum, the mass, and the decay width of the $W$ boson, respectively,
$d_{\mu\nu} = g_{\mu\nu} - q_{\mu}q_{\nu}/ m^2_{W}$ arises from the summation over the
polarizations of the $W$ boson, $P_{L}=({\bf 1}-\gamma^{5})/2$ is the
projection matrix, $\hat{\omega}_{t}=(1+\gamma_5 s\sura_{t})/2$ is the spin
projector for the top quark, $m_t$ and $m_b$ are the masses of
the top quark and the $b$ quark, respectively, and the lepton masses have
been dropped.

\begin{figure}
 \begin{center}
  \def\SCALE{0.40}
  \def\OFFSET{27pt}
  \begin{tabular}{c}
   \includegraphics[scale=\SCALE]{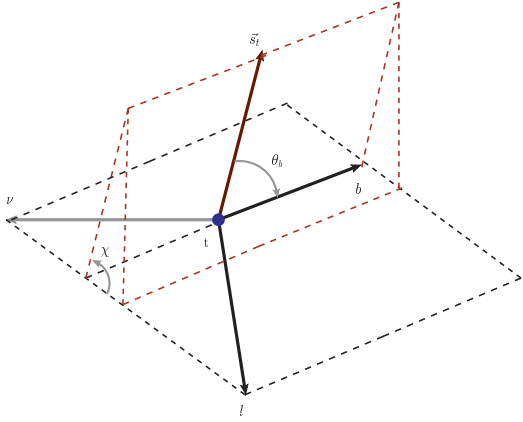}
  \end{tabular}
 \end{center}
 \begin{center}
 \caption{Angles of final-state particles defined in the rest
  frame of the top quark.}
 \label{fig4}
 \end{center}
\end{figure}

The three-body phase-space integral for the top quark decay
is given by
\begin{eqnarray}
\int d\mbox{PS}^{(3)}
 &=& \int \frac{1}{2E_{b}}\frac{d^3\vec{k}_{b}}{(2\pi)^3}
\frac{1}{2E_{\ell}}\frac{d^3 \vec{k}_{\ell}}{(2\pi)^3}
\frac{1}{2E_{\nu}}\frac{d^3 \vec{k}_{\nu}}{(2\pi)^3}
(2\pi)^4 \delta^4(k_t - k_{b} - k_{\ell} - k_{\nu}), \nonumber\\
 &=& \pi\int
     \frac{dE_{b} d\cos\theta_{b}}{2(2\pi)^2}
     \frac{dE_{\ell} d\chi}{2(2\pi)^3},\label{eq.def.phase.space}
\end{eqnarray}
with $E_i$ being the energy of the particle $i$, $i=b,\ell,\nu$.
The angle $\theta_{b}$ is the polar angle of the $b$ quark with
respect to the top quark spin as indicated in Fig.~\ref{fig4},
the azimuthal angle $\phi_{b}$ has been integrated out, and
$\chi$ denotes the angle between the $\vec{s}_{t}$-$\vec{k}_{b}$
plane and the $\vec{k}_{b}$-$\vec{k}_{\ell}$ plane.

We insert the identity
\begin{eqnarray}
\int dm_J^2dE_Jd^2\hat n_J\delta(m_J^2-m_b^2)\delta(E_J-E_b)
\delta^{(2)}(\hat{n}_J - \hat{n}_b) = 1,\label{i1}
\end{eqnarray}
into the differential decay width,
where $m_{J}$ ($E_{J}$) is the jet invariant mass (energy),
and the jet ($b$-quark) direction is defined
by ${\hat n}_J=\vec{k}_{J}/|\vec{k}_{J}|$ (${\hat n}_b=\vec{k}_{b}/|\vec{k}_{b}|$).
The three $\delta$-functions in Eq.~(\ref{i1}), together with the $b$-quark
phase space, are absorbed into the LO $b$-quark jet function
$J^{(0)}(m_J^2,E_J,R)$, which will be specified later.
We further factorize the fermion flow in
Eq.~(\ref{eq.M0sq}) by applying the Fierz transformation
\begin{eqnarray}
I_{ij}I_{lk}=\frac{1}{4}I_{ik}I_{lj}
+\frac{1}{4}(\gamma_5)_{ik}(\gamma_5)_{lj}
+\frac{1}{4}(\gamma_\alpha)_{ik}(\gamma^\alpha)_{lj}
+\frac{1}{4}(\gamma_5\gamma_\alpha)_{ik}(\gamma^\alpha\gamma_5)_{lj}
+\frac{1}{8}(\sigma_{\alpha\beta})_{ik}(\sigma^{\alpha\beta})_{lj},
\label{fi}
\end{eqnarray}
with $I$ being the identity matrix and $\sigma_{\alpha\beta}\equiv
i[\gamma_\alpha,\gamma_\beta]/2$. The vectors
\begin{eqnarray}
\xi_J=\frac{1}{\sqrt{2}}\left(1, -{\hat n}_{J}\right),\;\;\;\;
\bar\xi_J=\frac{1}{\sqrt{2}}\left(1,{\hat n}_{J}\right),
\end{eqnarray}
are introduced, which lie on the light cone and satisfy $\xi_J\cdot \bar\xi_J=1$.
The third term in Eq.~(\ref{fi}), especially the component
$(\not\xi_J)_{ik}(\not \bar\xi_J)_{lj}/4$, gives the leading-power contribution.
The matrix $(\not\xi_J/2)$ goes into the trace for the $b$-quark jet
function, and the matrix $(\not\bar\xi_J/2)$ goes into the trace for
the heavy-quark kernel. We also employ the identity
\begin{eqnarray}
I_{ij}I_{lk}&=&{1 \over
N_c}I_{ik}I_{lj}+2\sum_c(T^c)_{ik}(T^c)_{lj},\label{fierz}
\end{eqnarray}
to factorize the color flow, where $I_{ik}/N_c$ goes into the
$b$-quark jet function, and $I_{lj}$ goes into the heavy-quark kernel.
The fermion and color traces between the $b$-quark jet function
and the heavy-quark kernel are then separated.

\begin{center}
 \begin{figure}[htb]
  \includegraphics[scale=0.5]{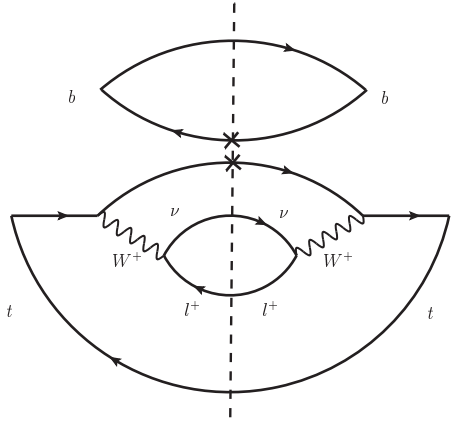}
  \caption{Factorization of the $b$-quark jet from
  the LO diagram for the decay width.} \label{fig3}
 \end{figure}
\end{center}

Figure~\ref{fig2} becomes Fig.~\ref{fig3}, where
the cross vertex stands for the insertion of the Gamma matrices
from the Fierz transformation.
The differential decay width is factorized at LO into
\begin{eqnarray}
d\Gamma^{(0)}(t\to b\ell\nu) &=&
\frac{\sqrt{2}\pi}{(2\pi)^5}\left(\frac{m_t}{2}\right)^3
d\cos\theta_J dm_J^2 \frac{x_Jdx_J}{\sqrt{1-4z_J/x_J^2}} dx_\ell
d\chi \\
&{}& \times H^{(0)}(k_t,k_\ell,k_J)J^{(0)}(m_J^2,E_J,R),\label{lof}
\end{eqnarray}
with the LO heavy-quark kernel
\begin{eqnarray}
H^{(0)}&=& \frac{1}{4}\frac{g^4|V_{tb}|^2}{(q^2-m^2_{W})^2
+m^2_{W}\Gamma^2_{W}} \nonumber\\
&{}& \times d_{\mu\nu}d_{\rho\sigma} \mbox{tr}\left[\gamma^{\nu}P_{L}{\not k}_{\ell}
           \gamma^{\sigma}P_{L}{\not k}_{\nu}\right]
\times \mbox{tr}\left[\gamma^{\mu}P_{L}\hat{w}_{t}({\not
k}_{t}+m_{t}) \gamma^{\rho}P_{L} \frac{1}{2} \not \bar\xi_J \right],\nonumber\\
&=& \frac{g^4|V_{tb}|^2}{2\sqrt{2}}
      \frac{1}{m_{t}}
     \frac{1}{x_{J}} f(x_{\ell}, x_{J}; z_{J})
(1+|\vec{P}|\cos\theta_{\ell}),\label{loh}
\end{eqnarray}
and the function
\begin{eqnarray}
f(x_{\ell}, x_{J}; z_{J}) &=&
\frac{ x_{\ell}(1 - x_{\ell} - z_{J}) }
       { (1 + z_{J} - x_{J} - \xi)^2 + \xi^2\eta^2 }.
\end{eqnarray}
In the above expression the dimensionless variables are defined as
\begin{eqnarray}
x_{J} = \frac{2E_J}{m_t},\hspace{0.3cm}
x_{\ell} = \frac{2E_{\ell}}{m_t},\hspace{0.3cm}
z_{J} = \frac{m^2_{J}}{m^2_{t}}, \hspace{0.3cm}
\xi = \frac{m^2_{W}}{m^2_{t}}, \hspace{0.3cm}
\eta=\frac{\Gamma_{W}}{m_{W}},
\end{eqnarray}
and the LO jet function $J^{(0)}$ is equal to the $\delta$-function,
$J^{(0)}=\delta(m_J^2-m_b^2)$. It is easy to see that Eq.~(\ref{loh})
is consistent with the fixed-order calculations in~\cite{TopdecayNLOlep1,TopdecayLO}.

\subsection{Factorization at NLO}

We then consider radiative corrections to the above LO decay width,
which involves two dramatically different scales, the top-quark mass
$m_t$ and the $b$-quark mass $m_b$. Since the $b$-quark mass will be
neglected eventually, $m_b$ and the QCD scale will not be differentiated
here. There are two leading infrared
regions for a loop momentum $l$, in which the radiative gluon is off-shell by
$l^2\sim O(m_b^2)$. The collinear region corresponds to $l$
collimated to $k_b$, namely, $k_b\cdot l\sim O(m_b^2)$. Another is
the soft region defined in the heavy-quark effective
theory, where the gluon momentum scales like $l^\mu \sim O(m_b)$ with
$k_b\cdot l\sim O(m_tm_b)$. Collinear gluons can be factorized from
the top quark decay amplitude into the $b$-quark jet function
$J(m_J^2,E_J,R)$ with the jet cone of radius $R$. Soft gluons, except
those emitted into the jet cone,
will be ignored as explained in the Introduction.
Hence, we focus on the factorization of the $b$-quark jet function below.

For the factorization at NLO, we first consider a virtual gluon
emitted by the top quark and attaching to the $b$ quark.
In the collinear region, where the loop momentum
$l$ carries a large component of $O(m_t)$ along the
$b$ quark momentum, we have the hierarchy
$k_t\cdot l \sim O(m_{t}^2)\gg l^2\sim O(m_b^2)\sim O(m_J^2)$.
It implies that the collinear gluon can be factorized into the
NLO virtual $b$-quark jet function $J^{(1)v}(m_J^2,E_J,R)$ by means of
the eikonal approximation, under which the virtual top
quark propagator is simplified into $\xi_J/\xi_J\cdot l$. It means
that the collinear gluon has been detached from the top quark
line, and collected by a Wilson line in the direction of $\xi_J$.
The separation of the
fermion flow and the color flow between the jet function and the heavy-quark
kernel also follows Eqs.~(\ref{fi}) and (\ref{fierz}), respectively.
Associating the three $\delta$-functions in Eq.~(\ref{i1}) with the
NLO jet function, the differential decay width with the above
virtual correction
is written, at leading power of $m_J/m_{t}$, as the convolution
\begin{eqnarray}
d\Gamma^{(1)v}(t\to b\ell\nu)
&=&\frac{\sqrt{2}\pi}{(2\pi)^5}\left(\frac{m_t}{2}\right)^3
d\cos\theta_J dm_J^2 \frac{x_Jdx_J}{\sqrt{1-4z_J/x_J^2}} dx_\ell
d\chi \nonumber\\
& &\times \left[ H^{(0)}(k_t,k_\ell,k_J)J^{(1)v}(m_J^2,E_J,R)\right.\nonumber\\
& &\left. \hspace{0.4cm} +H^{(1)v}(k_t,k_\ell,k_J)J^{(0)}(m_J^2,E_J,R)\right].
\label{hqv2}
\end{eqnarray}
The infrared-finite NLO virtual heavy-quark kernel $H^{(1)v}$ is defined
as the difference between the left-hand side and the first term on
the right-hand side via the above expression.

We then consider the QCD correction, where a real gluon is emitted by the
top quark and attaches to the $b$ quark. The hierarchy
$k_t\cdot l\sim O(m_{t}^2)\gg l^2=0$ ($k_t\cdot l\sim O(m_{t}m_b)\gg l^2=0$)
for a real gluon automatically holds in the collinear (soft) region, implying the
factorization of the infrared gluon into the
$b$-quark jet function through the eikonalization of the
top-quark propagator. In this case we insert the identity
\begin{eqnarray}
\int dm_J^2dk_J^0d^2\hat
n_J\delta(m_J^2-(k_b+l)^2)\delta(k_J^0-k_b^0-l^0) \delta^{(2)}(\hat
n_J-\hat n_{b+g})=1,\label{ire}
\end{eqnarray}
to define the NLO real $b$-quark jet function $J^{(1)r}(m_J^2,k_J^0,R)$,
where the $R$ dependence is introduced by requiring the
real gluon to be emitted within the jet cone of radius $R$, and
$\hat n_{b+g}$ denotes the direction of the total momentum of the $b$ quark
and the real gluon. The separation of the
fermion flow and the color flow is also achieved by applying Eqs.~(\ref{fi}) and
(\ref{fierz}). Combining the three $\delta$-functions in Eq.~(\ref{ire})
and the phase spaces for the $b$ quark and the gluon, together the normalization
factor $1/(k_J^0)^2$, we construct $J^{(1)r}(m_J^2,k_J^0,R)$.

The difference between the original diagram and the effective
diagram with the eikonal approximation gives the NLO real heavy-quark
kernel $H^{(1)r}$. For this difference, it is understood that the infrared
contribution in the top quark decay specifically associated
with the $b$ quark has been subtracted. The
top quark decay is then calculated as its final states are composed of a jet,
a real gluon, and a lepton pair at this order. Consequently, we insert
the identity in Eq.~(\ref{i1}), and associate the three
$\delta$-functions in Eq.~(\ref{i1}) with the LO $b$-quark jet function,
forming $J^{(0)}(m_J^2-2k_{J}\cdot l,E_J-l^0,R)$. Therefore, the differential
decay width is expressed as the convolution
\begin{eqnarray}
d\Gamma^{(1)r}(t\to b\ell\nu)
&=&\frac{\sqrt{2}\pi}{(2\pi)^5}\left(\frac{m_t}{2}\right)^3
d\cos\theta_J dm_J^2  \frac{x_Jdx_J}{\sqrt{1-4z_J/x_J^2}} dx_\ell
d\chi \nonumber\\
& &\times \left[ H^{(0)}(k_t,k_\ell,k_J)J^{(1)r}(m_J^2,E_J,R)\right.\nonumber\\
& &\left. \hspace{0.4cm} +\int\frac{d^3l}{2l^0(2\pi)^3}
H^{(1)r}(k_t,k_\ell,k_J,l,R)
J^{(0)}(m_J^2-2k_{J}\cdot l,
E_J-l^0,R)\right],\nn\\
\label{hqj2}
\end{eqnarray}
where the $R$ dependence of $H^{(1)r}$ arises from the subtraction of
$J^{(1)r}$ from the QCD diagram.
We should include the NLO diagrams into the complete heavy-quark kernel
$H^{(1)}$ with the radiative gluons being
exchanged among the top quarks. This category contains the
virtual diagrams, such as the self-energy correction to the top
quark, and the real diagrams, where gluons are exchanged between the top quarks on both
sides of the final-state cut. These
diagrams are also characterized by the scale $m_t$, i.e., dominated
by the contribution from $l$ not collimated to $k_J$.

\subsection{Factorization at All Orders}

To extend the jet factorization to all orders, we follow the procedures
outlined in \cite{Li01}: we first employ the collinear
replacement for the metric tensor of a gluon propagator \cite{Li01}
\begin{eqnarray}
g^{\alpha\beta}\to\frac{\xi_J^\alpha l^\beta}{\xi_J\cdot l},\label{metric}
\end{eqnarray}
where the vertex $\alpha$ ($\beta$) is located on the $b$-quark line
(all other lines in the QCD diagrams for the differential decay width).
It is easy to see that Eq.~(\ref{metric}) picks up the leading
collinear contribution. Applying the Ward identity
to the contractions of $l^\beta$ \cite{Li01}, it can be shown that
all the contractions are summed into the factorized form containing
the NLO jet functions $J^{(1)}$ derived
in the above subsections. The factor $\xi_J^\alpha/\xi_J\cdot l$
in Eq.~(\ref{metric}) corresponds to the Feynman rules for the top-quark
propagator under the eikonal approximation, namely, for the Wilson lines,
which are demanded by the gauge invariance of the jet function as a matrix element
of a nonlocal operator. Hence, a gluon collimated to the $b$ quark
is factorized from the $O(\alpha_s^{N+1})$ diagrams, leading to
$d\Gamma^{(N+1)}\approx d\Gamma^{(N)}\otimes J^{(1)}$. At last,
we establish the all-order factorization by induction \cite{Li01}
\begin{eqnarray}
d\Gamma(t\to b\ell\nu)
&=&\frac{\sqrt{2}\pi}{(2\pi)^5}\left(\frac{m_t}{2}\right)^3
d\cos\theta_J dm_J^2  \frac{x_Jdx_J}{\sqrt{1-4z_J/x_J^2}} dx_\ell
d\chi \nonumber\\
& &\times H(k_t,k_\ell,k_J,R)J(m_J^2,E_J,R).
\label{hqjt}
\end{eqnarray}
The above steps are basically the same as in \cite{Li01}, so the detail
will not be repeated in this work.

To verify that the jet function in Eq.~(\ref{hqjt}) absorbs the correct
logarithms, Eq.~(12) for virtual corrections and Eq.~(22) for real corrections
in the ``unresolved" region in \cite{TopdecayNLOhad} are expanded
in the small $z_b= m_b^2/m_t^2$ limit, and the large logarithms
$\ln^2 (z_b/x_b^2)$, with the energy fraction of the $b$ quark
$x_b= 2E_b/m_t$, are identified. The coefficients of these large
logarithms are then compared to those of $\ln\bar N$, with
$\bar N\equiv N\exp(\gamma_E)$, in the expansion of the Sudakov
exponent up to NLO (see Appendix A of \cite{Lijet}), where the variable
$N$ arises from the Mellin transformation in the jet mass space.
It is seen that both the double and single logarithms have the same
coefficients in the two references, except an additional single
logarithm proportional to $\ln(1/x_{\rm min}) - 9/4$ in \cite{TopdecayNLOhad}.
The dimensionless parameter $x_{\rm min}$ was introduced by slicing the
phase space of real corrections into the ``resolved" and ``unresolved"
regions, and the contribution from the former was obtained from a
numerical integration. We believe that the additional logarithm
proportional to $\ln(1/x_{\rm min}) - 9/4$ should be canceled,
once the ``resolved" contribution is also included in the comparison,
since the total result is independent of $x_{\rm min}$. An equivalent
postulation is that $\ln(1/x_{\rm min}) - 9/4$ can be made vanish by
choosing an appropriate value for the arbitrary $x_{\rm min}$.

As stated in the Introduction, the NLO corrections to the spin analyzing
power, being integrated over all final-state phase space, are not
important \cite{TopdecayNLOhad}. With the large logarithms having been
organized into the jet function, the remaining NLO corrections to the
heavy-quark kernel are expected to be negligible. Therefore, we will adopt
the LO heavy-quark kernel for numerical analysis below.
The differential decay width is then given by
\begin{eqnarray}
\frac{d\Gamma(t\to b l \nu)}
     {dm^2_{J}dx_{J}d\cos\theta_{J} dx_{\ell}}
 &=& 2\pi c^{\mbox{\tiny lep}} J(m^2_{J}, E_{J}, R)
     f(x_{\ell}, x_{J}; z_{J})
     \left( 1 + |\vec{P}| \cos\theta_{\ell} \right),\label{dou}
\end{eqnarray}
where the $\chi$ dependence has been integrated over, and the constant
$c^{\mbox{\tiny lep}}$ represents
\begin{eqnarray}
c^{\mbox{\tiny lep}}
= \frac{1}{4}\frac{1}{(2\pi)^4}G^2_{F}m^5_{t}\xi^2|V_{tb}|^2.
\end{eqnarray}


\section{RESULTS \label{results}}

We re-express Eq.~(\ref{dou}) in terms of the polar angle $\theta_J$
of the $b$-quark jet, and the angle $\theta_{J\ell}$
between the momenta of the charged lepton and of the $b$-quark jet,
\begin{eqnarray}
\frac{d\Gamma(t\to b \ell \nu)}
     {dm^2_{J}dx_{J}d\cos\theta_{J} dx_{\ell}}
 &=& 2\pi c^{\mbox{\tiny lep}} J(m^2_{J}, E_{J}, R)
     f(x_{\ell}, x_{J}; z_{J})
     \left( 1 + |\vec{P}| \cos\theta_{J\ell}\cos\theta_{J} \right),\label{dou1}
\end{eqnarray}
with
\begin{eqnarray}
\cos\theta_{J\ell}
&=& \frac{1}{\beta_{J}}
    \left[ 1 - \frac{2(x_{\ell} + x_{J} -1 -z_{J})}{x_{J}x_{\ell}}
    \right],
\end{eqnarray}
and $\beta_{J}=\sqrt{1-4z_{J}/x^2_{J}}$.
Further integrating over the energy fractions $x_{J}$ and
$x_{\ell}$, Eq.~(\ref{dou1}) gives
\begin{eqnarray}
\frac{1}{\Gamma}\frac{d\Gamma}{d\cos\theta_{J}dz_{J}}
 &=& \frac{1}{2A}
 \left[ a(z_J) + b(z_J)|\vec{P}|\cos\theta_{J}
\right], \label{eq.double.diff}
\end{eqnarray}
where $A$, $a(z_J)$, and $b(z_J)$ are written as
\begin{eqnarray}
A &=& \int_{0}^{1} dz_{J} ~a(z_{J}),
\hspace{0.5cm}\Gamma=4\pi c^{\mbox{\tiny lep}}A ,\nn\\
a(z_{J}) &=& \int_{x_{J\min}}^{x_{J\max}} dx_{J}~ J(z_J, x_J,R) F_{a}(z_J, x_J),\nn\\
b(z_{J}) &=& \int_{x_{J\min}}^{x_{J\max}} dx_{J}~ J(z_J, x_J,R) F_{b}(z_J, x_J),
\label{ab}
\end{eqnarray}
with $x_{J\min}=2\sqrt{z_{J}}$, $x_{J\max}=1 + z_{J}$, and
\begin{eqnarray}
F_{a}
&=& \frac{x_{J}\beta_{J}}{[1 + z_{J} - x_{J} - \xi]^2+(\xi\eta)^2}
    \left[ - \frac{1}{3}x^2_{J} + \frac{1+z_{J}}{2}x_{J} - \frac{2z_{J}}{3}
    \right], \nn\\
F_{b}
&=& \frac{1}{[1 + z_{J} - x_{J}    - \xi]^2+(\xi\eta)^2}
    \left[ - \frac{1}{3}x^3_{J}    + \frac{1+3z_{J}}{6}x^2_{J}
		   + \frac{4z_{J}}{3}x_{J} - \frac{2}{3}z_{J}(1 + 3z_{J})
    \right].
\end{eqnarray}
The $R$ dependencies of $a(z_{J})$ and $b(z_{J})$ in Eq.~(\ref{ab}) are implicit.

The spin analyzing power for a final state $i$ in a polarized top quark decay
can be defined via the average of $\cos\theta_i$
\begin{eqnarray}
 \langle \cos\theta_{i} \rangle
\equiv \frac{1}{\Gamma}\int d\cos\theta_{i}
\cos\theta_{i}
  \frac{d\Gamma}{d\cos\theta_{i}},
\end{eqnarray}
with which the usual expression in Eq.~(\ref{par})
leads to $\kappa_{i} = 3 \langle \cos\theta_{i} \rangle$. Following
the same reasoning, we can study the spin analyzing power for the $b$-quark jet
with a specified invariant mass,
\begin{eqnarray}
\kappa_{J}(z_J)\equiv 3 \langle \cos\theta_J \rangle
&=& \frac{3}{\Gamma}\int d\cos\theta_{J}
\cos\theta_{J}
  \frac{d\Gamma}{d\cos\theta_{J}dz_{J}}
= \frac{b(z_J)}{a(z_J)}. \label{eq.avecos.zj.distribution}
\end{eqnarray}
That is, the function $a(z_J)$ is related to the
decay width for producing a $b$-quark jet of mass $m_J$, and the function
$b(z_J)$ is related to angular distribution of the $b$-quark jet.
In this paper, we will evaluate $a(z_J)$, $b(z_J)$, and $\kappa_{J}(z_J)$, and
discuss the effect on the spin analyzing power from the inclusion of
the jet function.

It has been observed that the $b$ quark mass is negligible
in the analysis of the polarized top quark decay according to \cite{TopdecayNLOhad},
implying the approximation of the $b$-quark jet
function by the light-quark jet function.
Moreover, it was noticed that the differential
cross sections for jet production scale with the ratio
$y\equiv m_{J}/(RE_{J})$ \cite{Lijet}.
To simplify the numerical calculation,
we are allowed to parameterize the light-quark jet function,
\begin{eqnarray}
J(y) =
\begin{cases}
 ~0 \hspace{6.3cm}\mbox{for} \hspace{0.2cm} y \le y_{0} ,\\
 ~\frac{J_0}{(RE_J)^2}\left(y - y_{0}\right)^{k_1}
  \exp\left(- k_2 y^2\right)
  \hspace{1.2cm}\mbox{for} \hspace{0.2cm} y > y_{0},\label{para}
\end{cases}
\end{eqnarray}
where $1/(R E_{J})^2$ fixes the dimension of the jet function.
The normalization constant $J_{0}$, determined by
$\int dm^2_{J} J(m^2_{J}, E_{J}, R) = 1$, does not
affect the evaluation of the spin analyzing power as indicated by
Eq.~(\ref{eq.avecos.zj.distribution}).
Fitting Eq.~(\ref{para}) to Figs.~4 and 5 in \cite{Lijet}, which have involved
non-perturbative contribution from the region with large Mellin moments, we obtain
$J_{0}=491.4$, $y_{0}=0.05$ and $k_1=0.6$, and $k_2=75$.
It should be stressed that the parameterized jet function must have some deviation
from the exact one, and the soft function has been neglected here. Hence, results
presented below can be regarded as being derived from a theoretically well
motivated phenomenological study.

\begin{figure}
  \begin{center}
    \def\SCALE{0.65}
    \def\OFFSET{27pt}
    \begin{tabular}{cc}
      \includegraphics[scale=\SCALE]{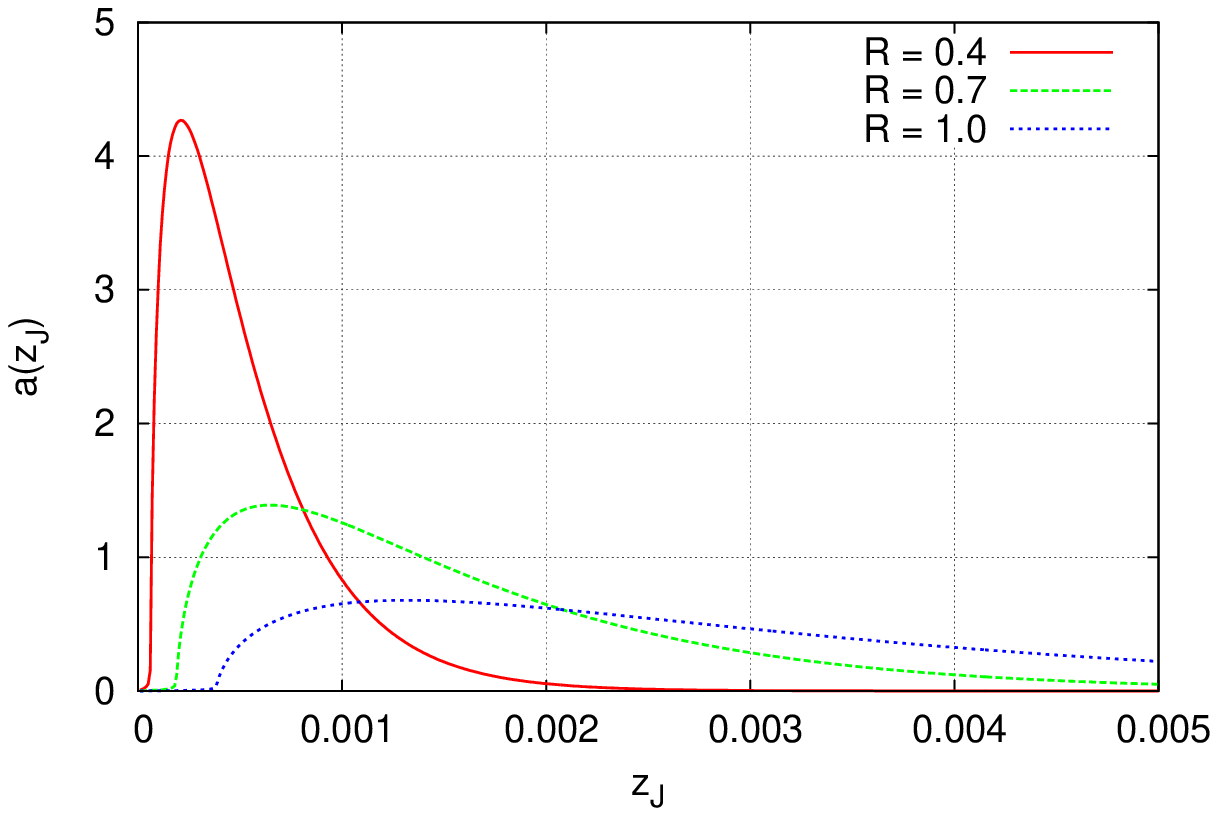} &
      \includegraphics[scale=\SCALE]{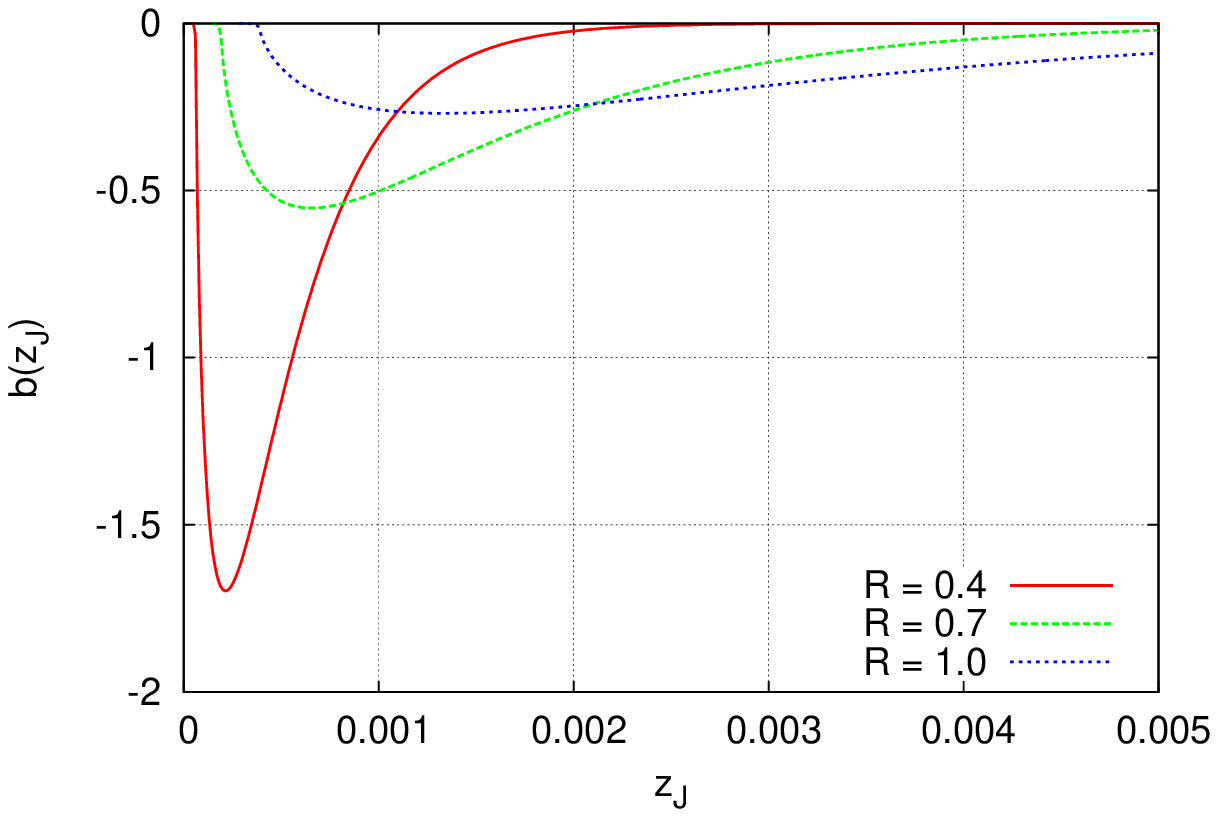} \\
      \hspace{\OFFSET} (a)
    & \hspace{\OFFSET} (b)
    \end{tabular}
    \caption{Dependencies of (a) $a(z_J)$ and of (b) $b(z_J)$ on $z_J$.}
    \label{fig56}
  \end{center}
\end{figure}
We adopt the following inputs~\cite{PDG.data}
\begin{eqnarray}
 m_{t} = 173.5~\mbox{GeV}, \hspace{0.5cm}
 m_{W}  = 80.39~\mbox{GeV}, \hspace{0.5cm}
 \Gamma_{W} = 2.085~\mbox{GeV},
\end{eqnarray}
and vary the jet radius to examine its influence by choosing $R=0.4$, $0.7$, and $1.0$.
The jet mass dependencies of $a(z_J)$ and $b(z_J)$ with three different values of $R$
are displayed in Figs.~\ref{fig56}(a) and \ref{fig56}(b), respectively.
The peaks arise from the
convolutions of the jet function with the kernels $F_{a}$ and $F_{b}$ that
contain the Breit-Wigner structure in the $W$-boson propagator.
Increasing the jet radius, the peaks of $a(z_J)$ and $b(z_J)$ shift
toward the large mass region, and both distributions become broader,
because more infrared radiations are included.
The peak positions $z_{J}^{\mbox{\tiny peak}}$ ($m_{J}^{\mbox{\tiny peak}}$),
almost identical for $a(z_J)$ and $b(z_J)$,
are found to be 2.1 $\times 10^{-4}$ (2.5 GeV), 6.5 $\times 10^{-4}$ (4.4 GeV),
and 1.3 $\times 10^{-3}$ (6.3 GeV) with $R=0.4$, 0.7, and 1.0, respectively.
They are of the same order of magnitude as the
mass ratio between the $b$ quark and the top quark,
$z_{b}= 5.8 \times 10^{-4}$ for $m_{b}=4.18~\mbox{GeV}$.

\begin{figure}
  \begin{center}
    \def\SCALE{0.65}
    \def\OFFSET{27pt}
    \begin{tabular}{cc}
      \includegraphics[scale=\SCALE]{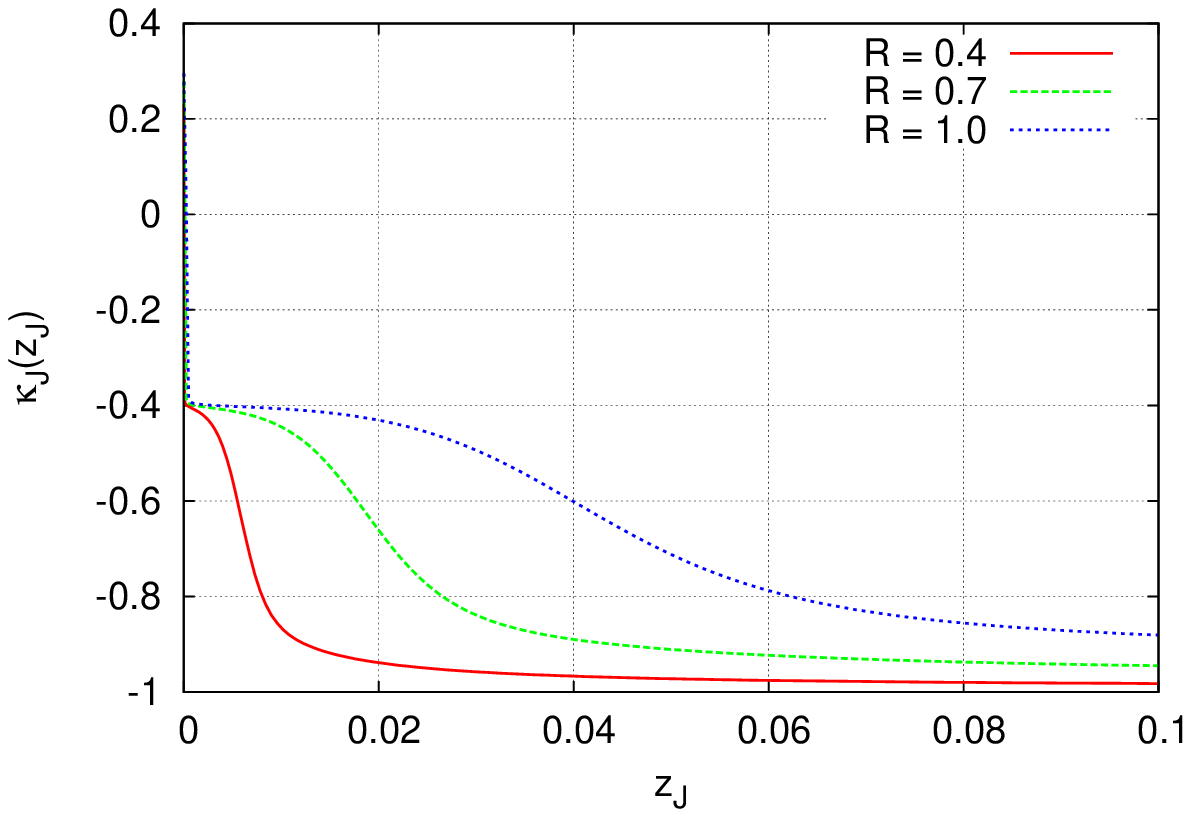} &
      \includegraphics[scale=\SCALE]{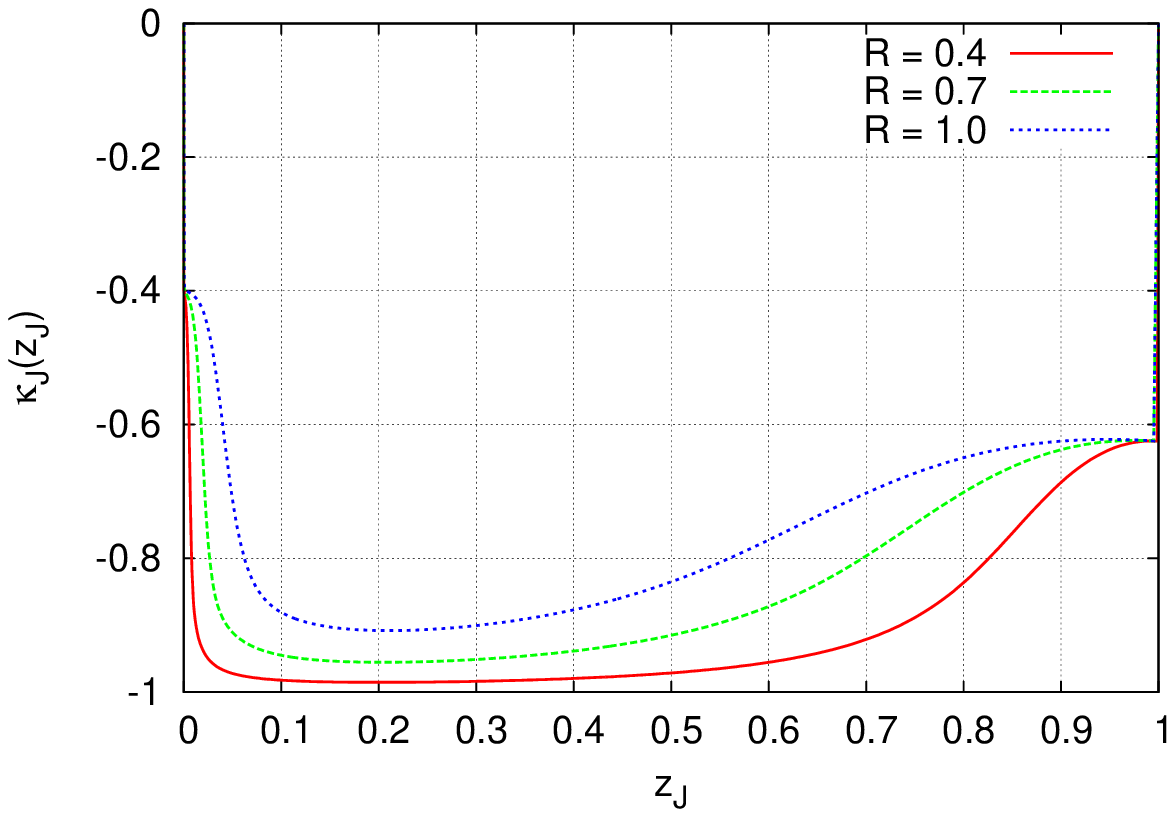} \\
      \hspace{\OFFSET} (a) &
      \hspace{\OFFSET} (b)
    \end{tabular}
    \caption{Dependence of $\kappa_J$ on
  $z_J$ in the ranges of (a) $0< z_J <0.1$ and of (b) $0 < z_J <1$.
  The ratios $z_J=0.001$, $0.005$, $0.01$, $0.05$,
  $0.1$, $0.5$, and $1.0$ correspond to the jet masses $m_J=5.5~\mbox{GeV}$,
  $12.3~\mbox{GeV}$, $17.4~\mbox{GeV}$, $38.8~\mbox{GeV}$,
  $54.9~\mbox{GeV}$, $122.7~\mbox{GeV}$, and $173.5~\mbox{GeV}$, respectively.}
    \label{fig78}
  \end{center}
\end{figure}

The dependence of the spin analyzing power $\kappa_J(z_J)=b(z_J)/a(z_J)$
on the jet mass is shown in Fig.~\ref{fig78}, with
its behavior at small $z_J$ being highlighted  in Fig.~\ref{fig78}(a).
It indicates that all curves agree with $\kappa_J\approx -0.4$
obtained in the literature at $z_J\approx 0$. The spin analyzing
power then increases rapidly with the jet mass, stabilizes
around $\kappa_J=-0.9\sim -1$ (depending on the jet radius)
in a wide mass region $0.1 <z_J < 0.5$,
and slowly decreases at large $z_J$. At last, the spin analyzing
power vanishes quickly at $z_J=1$ as expected, which corresponds to the
kinematic boundary. Another important feature is that $\kappa_J(z_J)$ is
insensitive to the choice of the jet radius in the middle mass region.
Although the jet production is dominated by the mass region around $m_b$,
it is still possible to determine the spin of a polarized top quark by
measuring the angular distribution of the $b$-quark jet with
a bit higher mass. In particular, the broader mass distribution at a larger jet radius
warrants this possibility against lower event rates.
We compare $\kappa_J$ in Eq.~(\ref{eq.avecos.zj.distribution}) based on the
jet factorization with the usual spin analyzing power $\kappa_b$ based on
the parton model. Replacing the jet function by its LO expression
$\delta(m_J^2-m_b^2)$, namely, regarding the $b$ quark as a parton,
we get the spin analyzing power $-0.401$. This value is close to the fixed-order
(LO or NLO) results in~\cite{TopdecayNLOlep1,TopdecayNLOlep2} for
semi-leptonic decays and in~\cite{TopdecayNLOhad} for hadronic three-body
decays. Integrating $a(z_J)$ and $b(z_J)$ over the jet mass, our formalism
reduces to the usual spin analyzing power, giving
$-0.407$, $-0.402$, and $-0.400$ for $R=0.4$, 0.7, and 1.0, respectively.
The above checks confirm the consistency of our formalism.

\begin{figure}
  \begin{center}
    \def\SCALE{0.65}
    \def\OFFSET{27pt}
    \begin{tabular}{cc}
      \includegraphics[scale=\SCALE]{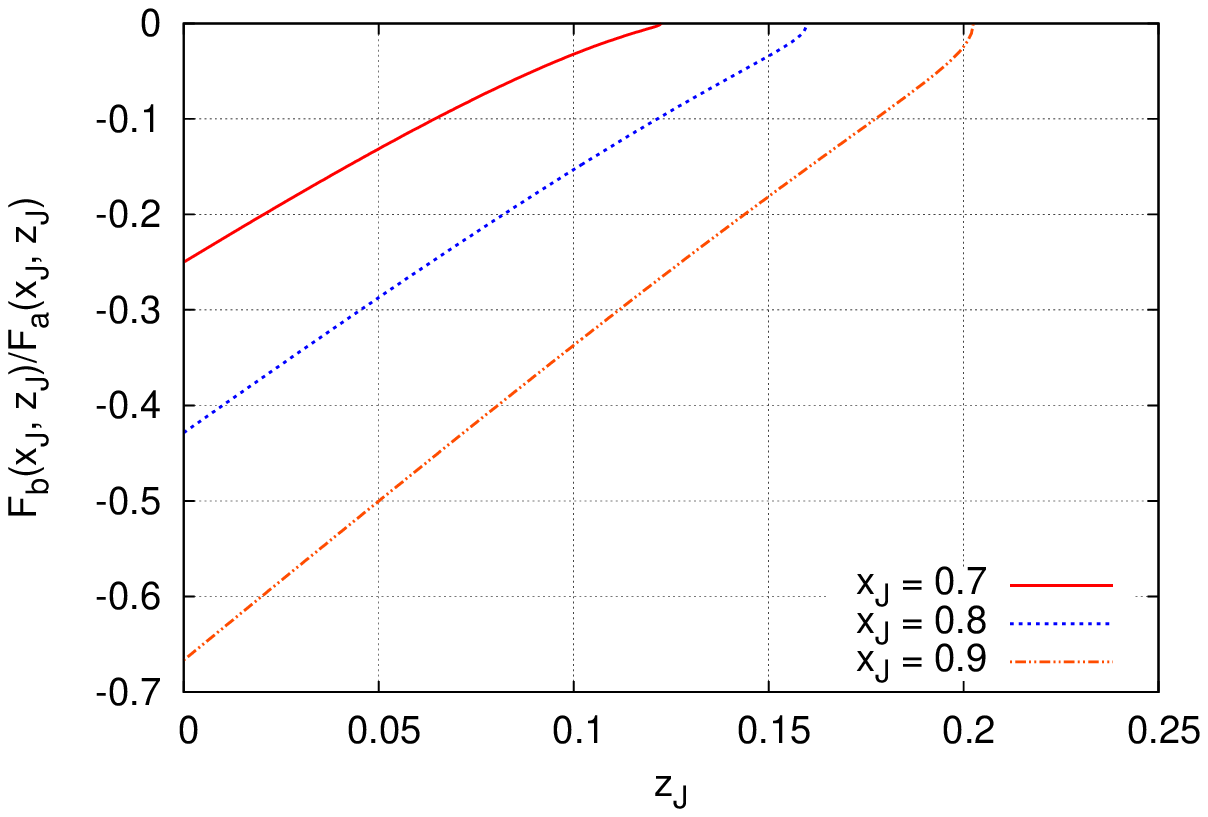} &
      \includegraphics[scale=\SCALE]{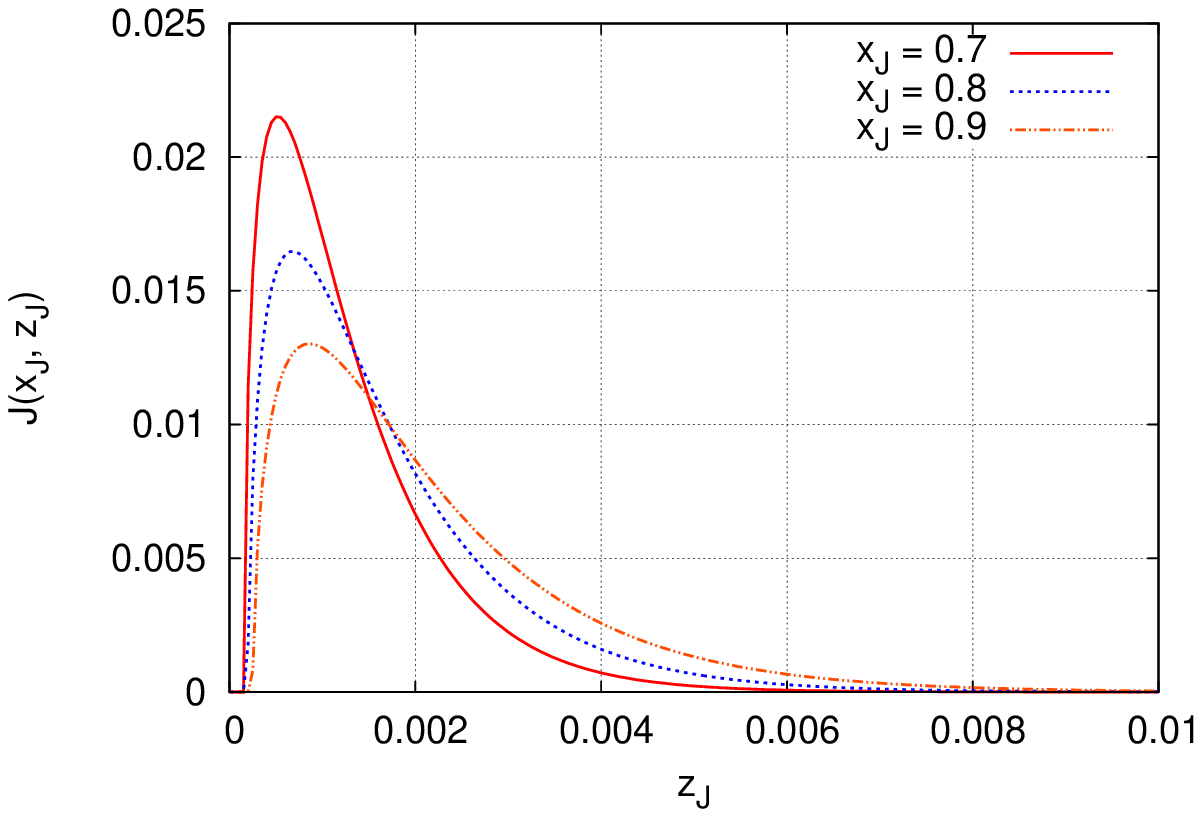} \\
      \hspace{\OFFSET} (a)
    & \hspace{\OFFSET} (b)
    \end{tabular}
    \caption{Dependencies of (a) the ratio $F_{b}/ F_{a}$ and of
    (b) the jet function $J$ with $R=0.7$ on $z_J$ for $x_J=0.7,0.8$, and $0.9$.
    The peak positions for the
   jet function are $5.0\times 10^{-4}(3.9~\mbox{GeV})$, $7.0\times
   10^{-4}(4.6~\mbox{GeV})$, $8.5\times 10^{-4}(5.1~\mbox{GeV})$,
   in $z_J$ ($m_J$) for $x_J=0.7$, $0.8$, and $0.9$, respectively.}
    \label{fig910}
  \end{center}
\end{figure}

The possible enhancement of the spin analyzing power by considering
the jet function is explained as follows.
We first investigate the jet-mass dependence of the ratio
$F_{b}/ F_{a}$ in Fig.~\ref{fig910}(a), and observe
that $|F_{b}/F_{a}|$ for a given $x_J$ decreases with $z_J$ from an
initial value at $z_J=0$. Plotting
the $x_J$ dependence of $F_{a}$ and $F_{b}$, it is seen that
the curve exhibits a peak at $x_J\approx 0.8$ for a wide range of $z_J$,
namely, $x_J = 0.8$ is a typical jet energy
fraction in a top quark decay. It
accounts for the usual spin analyzing power derived in
the literature, because the initial value is about
$F_{b}/F_{a} \approx -0.4$ as $x_J = 0.8$. The initial value increases
with $x_J$, and reaches around $-0.7$ as $x_J = 0.9$. To enhance
the spin analyzing power, the large $x_J$ region must be selected.
We then display the behavior of the $b$-quark
jet function with $R=0.7$ in Fig.~\ref{fig910}(b), where it
exhibits weaker suppression at $z_J$ slightly higher than zero,
say, $z_J > 0.002$, as $x_J$ increases. That is, Fig.~\ref{fig910}(b)
shows the known broadening of a jet in the mass distribution with the jet energy.

Since $F_{b}/F_{a}$ with $x_J>0.8$ gets more weight as convoluted with
the jet function, the spin analyzing power grows with $z_J$. To confirm that
the jet energy dependence plays an important role, we test a model jet function,
which is energy independent, $J(m^2_{J})=\theta(m_0^2-m_J^2)/m_0^2$. Inserting this
simple model into Eq.~(\ref{ab}), we obtain the spin analyzing power
that always decreases with $z_J$. It is worthwhile to test experimentally
the enhancement of the spin analyzing power through
the inclusion of jet dynamics.

\section{Conclusion \label{conclusion}}

In this paper we have performed pQCD factorization of the
$b$-quark jet function from the semi-leptonic decay of a polarized top quark
by means of the eikonal approximation and the Ward identity.
The resultant formula is expressed as the convolution of the infrared-finite
heavy-quark kernel with the jet function, which is dominated by infrared radiations.
Adopting the LO heavy-quark kernel and parameterizing the jet
function from QCD resummation, we have predicted the dependence of the spin
analyzing power associated with the $b$-quark jet
on its invariant mass. It has been verified that our formalism
reproduces the fixed-order results by integrating over the jet mass.
Our investigation indicated that the spin
analyzing power could be possibly enhanced by a factor 2 as measuring the angular
distribution of the $b$-quark jet with a specified invariant mass. The
mechanism responsible for the enhancement has been elaborated, which is attributed
to the broadening of a jet in the mass distribution with the jet energy. This
broadening puts more weight on the contribution
from the region with higher jet energy, that intends to give a larger spin analyzing
power. The above observation is rather insensitive to the choice of the jet radius,
and worth of experimental confrontation in view of deeper understanding of
the effect from including jet dynamics into the polarized top quark decay.

As future works, NLO corrections to the heavy-quark
kernel need to be taken into account in order to improve
the precision of our predictions. At this level, the dependence of the top quark
mass definition on renormalization schemes (for instance, the $\overline{\rm MS}$
mass \cite{msbarmass1,msbarmass2} or the pole mass), becomes an essential issue.
It is possible to extend the current formalism to hadronic decays of a
polarized top quark, whose analysis is more complicated.
Another important direction is to develop the formalism for
a boosted top quark at LHC \cite{Boost1,Boost2}, that mainly decays into
a single jet. Substructures for a polarized top quark jet, which
improve the extraction of the information on the top quark polarization,
can be calculated.

\section*{Acknowledgment}
We thank Z.G. Si for useful discussions.
This work was supported in part by the National Science Council of
R.O.C. under Grant No. NSC-101-2112-M-001-006-MY3, and by the National
Center for Theoretical Sciences of R.O.C..


\begin{thebibliography}{99}

\bibitem{Topreview1} F.-P. Schilling,
	Int.~J.~Mod.~A {\bf 27}, 1230016 (2012).

\bibitem{Topreview2} W.~Bernreuther,
	J.~Phys.~G.~Nucl.~Part.~Phys.~{\bf 35}, 083001 (2008).
	
\bibitem{TopreviewNLO} J.~M.~Campbell and R.~Keith Ellis,
		arXiv:1204.1513~[hep-ph].

\bibitem{New}
X. Gong, Z.-G. Si, S. Yang, and Y.-j. Zheng, arXiv:1210.7822 [hep-ph];
S.M. Troshin and N.E. Tyurin, arXiv:1210.0394 [hep-ph];
S. Fajfer, J.F. Kamenik, and B. Melic, arXiv:1205.0264 [hep-ph].

\bibitem{TopdecayWNLO} M.~Fischer, S.~Groote, J.~G.~K\"orner, and M.~C.~Mauser,
	Phys. Rev. D {\bf 65}, 054036 (2002).

\bibitem{TopdecayNLOlep1} M.~Jezabek and J.~H.~K\"uhn,
	 Nucl.~Phys.~B {\bf 320}, 20 (1989);

\bibitem{TopdecayNLOlep2}	 A.~Czarnecki, M.~Jezabek, and J.~H.~K\"uhn,
	 Nucl.~Phys.~{\bf B351}, 70 (1991).

\bibitem{TopdecayNLOhad}	 A.~Brandenburg, Z.~G.~Si, and P.~Uwer,
	 Phys.~Lett.~B {\bf 539}, 235 (2002).

\bibitem{less.energetic} M.~Jezabek,
	Nucl. Phys. (Proc.~Supple.) {\bf 37B}, 197 (1994).

\bibitem{Factorization1}
	J.~C.~Collins, D.~E.~Soper, and G.~Sterman,
	Nucl. Phys. {\bf 261}, 172 (1985); {\bf 308}, 833 (1988).

\bibitem{Factorization2}
	G.~T.~Bodwin, Phys. Rev. D {\bf 31}, 2616 (1985); {\bf 34}, 3932 (1986).	

\bibitem{Factorization3} J.~C.~Collins and G.~Sterman,
	Nucl.~Phys.~ {\bf 185}, 172 (1981).

\bibitem{Li01} H.-n. Li, Phys. Rev. D {\bf 64}, 014019 (2001);
M. Nagashima and H.-n. Li, Eur. Phys. J. C {\bf 40}, 395 (2005).

\bibitem{Collins.review} J.~C.~Collins,
	Adv.~Ser.~Direct.~High Energy Phys.~{\bf 5}, 573 (1989)
	[hep-ph/0312336].

\bibitem{Factorization.old1} A.~H.~Mueller,
	Phys. Rev. D {\bf 20}, 2037 (1979).

\bibitem{Factorization.old2} J.~C.~Collins and D.~E.~Soper,
	Nucl. Phys. {\bf 193}, 381 (1981).

\bibitem{Factorization.old3} A.~Sen,
	Phys. Rev. D {\bf 24}, 3281 (1981).	

\bibitem{Factorization.review} H.~Contopanagos, E.~Laenen, and G.~Sterman,
	Nucl. Phys. {\bf 484}, 303 (1997).

\bibitem{Lijetl} H.-n. Li, Z. Li, and C.-P. Yuan,
Phys. Rev. Lett. {\bf 107}, 152001 (2011).

\bibitem{Lijet} H.-n. Li, Z. Li, and C.-P. Yuan,
	 arXiv:1206.1344.

\bibitem{TopdecayLO} K.~Fujikawa, Prog.~Theo.~Phys.~{\bf 61}, 1186 (1979);
	             J.~H.~K\"uhn, Acta.~Phys.~Pol.~{\bf B12}, 374 (1981);
	  J.~H.~K\"uhn and K.~H.~Streng, Nucl~Phys.~{\bf B198}, 71 (1982).

\bibitem{PDG.data} J.~Beringer {\sl et al.} (Particle Data Group),
	 Phys. Rev. D{\bf 86}, 010001 (2012).

\bibitem{Boost1} J.~Shelton,
	 Phys. Rev. D {\bf 79}, 014032 (2009).

\bibitem{Boost2} D.~Krohn, J.~Shelton, and L.~T. Wang,
	 JHEP {\bf 07}, 041 (2010).

\bibitem{msbarmass1} U.~Langenfeld, S.~Moch, and P.~Uwer,
	 Phys. Rev. D {\bf 80}, 054009 (2009).
	
\bibitem{msbarmass2} V.~Ahrens, A.~Ferroglia, M.~Neubert, B.~D.~Pecjak, and L.~L.~Yang,
	JHEP {\bf 09}, 097 (2010).
	
\end{thebibliography}


\end{document}